\documentclass[11pt,twoside]{article}

\usepackage{asp2006}
\usepackage{epsf}
\usepackage{lscape}

\begin{document}
\title{Anisotropic Outflows and Enrichment of the Intergalactic Medium}   
\author{Matthew M. Pieri, Hugo Martel, C\'edric Grenon}   

\affil{D\'epartement de physique, de g\'enie physique et
d'optique, Universit\'e Laval, Qu\'ebec, QC, G1K 7P4, Canada}    

\begin{abstract} 
We have developed an analytical model for the evolution
of anisotropic galactic outflows. These
outflows follow the path of least resistance, and thus
travel preferentially into low-density regions, away from cosmological
structures where galaxies form. 
We show that
aniso\-tropic outflows can significantly enrich low-density
systems, while reducing the enrichment of overdense regions. 
\end{abstract}


\section{Introduction}

Galactic outflows play an important role in the evolution of galaxies
and the Intergalactic Medium (IGM). Supernova explosions in galaxies
create galactic winds, which deposit energy and metal-enriched gas into the
IGM. Simulations of explosions in a single
object reveal that outflows tend
to be highly anisotropic, with the energy and metal-enriched
gas being channeled along the direction of least resistance
(Martel \& Shapiro 2001).
Several observations also support the existence of anisotropic
outflows both directly
(e.g. Bland \& Tully 1988; Veilleux \& Rupke 2002) and indirectly (Pieri \& Haehnelt 2004;
Pieri, Schaye, \& Aguirre 2006).

\begin{figure}
\hskip0.6in
\plotfiddle{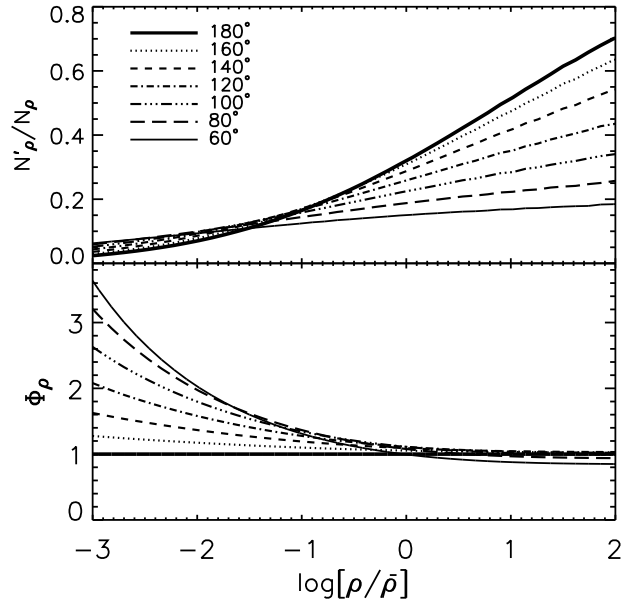}{3.in}{90}{48}{48}{230}{-20}
\caption{{\it Top}: fraction of enriched grid points 
$N^\prime_\rho/N_\rho$ in the simulation volume at $z=2$ 
as a function of gas overdensity $\rho/\bar\rho$
for varying opening angle. 
{\it Bottom}: number of enriched points below an overdensity threshold,
relative to the isotropic case.}
\end{figure}

We represent outflows as two bi-polar spherical cones traveling 
along a path of least resistance,
with an opening angle, $\alpha$, that can vary from $180^\circ$ 
(isotropic outflows) to $\sim60^\circ$. The direction of these 
outflows is determined by the path of least resistance on the halo 
collapse scale. We describe the expansion of the outflows using the
formalism of Tegmark, Silk, \& Evrard (1993), modified for anisotropy.
To implement this outflow model in a cosmological simulation,
we use the Monte Carlo method of
Scannapieco \& Broadhurst (2001). 

We track the formation of $\sim 20000$ galaxies in a
comoving cubic volume of size $12h^{-1}{\rm Mpc}$, in
a $\Lambda$CDM universe. When these outflows strike halos in the
process of collapsing, the outflows may ram-presure-strip them of their 
baryons. Where the halo is not stripped, it is enriched by the metals 
carried by these outflows.  The deposition of metals modify the cooling 
time of gas in halos and so the redshift of galaxy formation.

\section{Results}

To investigate the nature of the regions enriched in metals by outflows,
we calculate the gas density and metallicity inside the computational
volume, on a $512^3$ grid (Pieri, Martel \& Grenon 2006). 
The top panel of Figure~1 shows the number of grid points $N^\prime_\rho$
enriched at a given gas overdensity $\rho/{\bar \rho}$, as a fraction of the total number 
of grid points $N_\rho$ at that density,
for a range of different opening angles; this is 
effectively the probability 
of enriching a systems of a given density. The 
impact of galactic outflows on overdense systems is dramatically reduced for 
increasingly anisotropic outflows, while 
the probability of enriching low-density (around mean density of the 
Universe or lower) systems increases.

The bottom panel shows $\Phi_\rho = N^\prime_{\rho^\prime<\rho}/
N^\prime_{\rho^\prime<\rho,180^\circ}$,
the number of enriched grid points with a overdensity 
below a given threshold $\rho$,
relative to the isotropic case.
This highlights the significant impact on 
the enrichment of underdense systems by anisotropic outflows. 
This can lead to an increase in the enriched volume of underdense systems 
of $10\%$ (where $\alpha=100^\circ- 120^\circ$) and an increase of $40\%$
in systems below $\rho/{\bar \rho}=0.1$ (where $\alpha=80^\circ- 100^\circ$) 
compared to isotropic outflows.



\end{document}